\documentclass[twocolumn,reprint,aps,prb, floatfix,amsmath,amssymb,superscriptaddress]{revtex4-2}
\usepackage[utf8]{inputenc}
\usepackage{graphicx}
\usepackage{bm}
\usepackage{braket}
\usepackage{xcolor} 
\usepackage{subfigure}     
\usepackage{amsthm}
\usepackage{hyperref}
\usepackage{makecell}

\newcommand{\cyl}{cyl}
\newcommand{\stp}{stp}
\newcommand{\PBC}{{\textsl{\tiny PBC}}}
\newcommand{\UHP}{{\textsl{\tiny UHP}}}
\begin{document}

\title{Universal information of critical quantum spin chains from wavefunction overlaps}
\date{\today}
\author{Yijian Zou}
\affiliation{Sandbox@Alphabet, Mountain View, CA 94043, USA}

\begin{abstract}
Given a critical quantum spin chain, we show how universal information about its quantum critical point can be extracted from wavefunction overlaps. More specifically, we consider overlap between low-energy eigenstates of the spin chain Hamiltonian with different boundary conditions, namely periodic boundary conditions and open boundary conditions. We show that such overlaps decay polynomially with the system size, where the exponent only depends on the central charge. Furthermore, the bulk-to-boundary operator product expansion (OPE) coefficients can be extracted from the overlaps involving excited states. We illustrate the proposal with the Ising model and the three-state Potts model.
\end{abstract}

\maketitle
\section{Introduction}
Universality is one of the most remarkable properties of critical phenomena. At long distances, different microscopic models can be characterized by the same universal data \cite{wilson_renormalization_1974,wilson_renormalization_1975}. In 1+1 dimensions, a critical quantum spin chain is usually described by a conformal field theory (CFT) \cite{belavin_infinite_1984,friedan_conformal_1984}, which is in turn completely characterized by the conformal data. Given a critical quantum spin chain, it is highly nontrivial to extract the conformal data of the underlying CFT. 

For 1+1 dimensional CFTs, several geometries are relevant, including the plane and the upper half plane. For a CFT on the plane, the conformal data includes a set of primary operators $\phi_{\alpha}$ with scaling dimensions $\Delta_{\alpha}$ and conformal spins $s_{\alpha}$, the central charge $c$, and operator product expansion (OPE) coefficients which involve three primary operators \cite{sonoda_sewing_1988}. For a CFT on the upper half plane, one further needs to specify the boundary condition, which we label as $a$. The conformal data includes a set of boundary primary operators $\psi^{a}_{\beta}$ with scaling dimensions $h^{a}_{\beta}$ \cite{cardy_conformal_1984_2}. In order to determine correlation functions involving both bulk and boundary operators, one additionally needs the boundary-to-boundary OPE coefficients which involve three boundary operators, and bulk-to-boundary OPE coefficients $C^{a}_{\alpha\beta}$ which involve a bulk operator and a boundary operator \cite{lewellen_sewing_1992}. For rational CFTs, the bulk and boundary conformal data are closely related \cite{lewellen_sewing_1992, cardy_bulk_1991, cardy_boundary_1989}, and they can be derived systematically from the Moore-Seiberg data \cite{moore_classical_1989} of the CFT. However, much less is known for generic, non-rational CFTs. Therefore, it is desirable to compute both the bulk and boundary conformal data with a numerical method.

One strategy to extract the conformal data from a quantum spin chain is based on the operator-state correspondence \cite{francesco_2012,henkel_1999}, pioneered by Cardy and others in the 80s. Such a strategy starts with diagonalization of the low-energy spectrum of the critical quantum spin chain on a finite number of sites. The scaling dimensions of scaling operators are then obtained by the energy spectrum of the spin chain. 
The approach applies for both spin chains with periodic boundary conditions (PBC) \cite{cardy_operator_1986,affleck_universal_1986,blote_1986} and spin chains with open ends \cite{cardy_effect_1986}. For a critical quantum spin chain with PBC, a low-energy eigenstate corresponds to a bulk scaling operator, where one can identify the states $|\phi^{\PBC}_{\alpha}\rangle$ that correspond to primary operators $\phi_{\alpha}$. For a critical quantum spin chain with open ends, if both ends have the same boundary condition labelled by $a$, then a low-energy state is in one-to-one correspondence with a boundary operator, where one can also identify the states $|\psi^{a}_{\beta}\rangle$ that correspond to boundary primary operators $\psi^{a}_{\beta}$. In order to extract OPE coefficients, however, one has to go beyond the spectrum alone \cite{cardy_operator_1986} and identify lattice realizations of CFT operators. This is hard to obtain if the model cannot be exactly solved (see however \cite{zou_conformal_2020}).

In this work, we show that conformal data, especially the bulk-to-boundary OPE coefficients, can already be extracted from the eigenstates without computing a lattice realization of CFT operators. Specifically, we consider the overlaps $\langle \psi^{a}_{\beta} |\phi^{\PBC}_{\alpha} \rangle$, which involve eigenstates of the critical quantum spin chain with the same number of spins and different boundary conditions. Our main result is that (i) the overlap between ground states decay polynomially with the system size $N$,
\begin{equation}
\label{eq:main_result1}
    \bra{I^{a}} I^{\PBC}\rangle = \mathcal{N}_a N^{-c/16},
\end{equation}
where $\mathcal{N}_a$ is a non-universal constant depending on lattice realization of the boundary condition, and $c$ is the central charge, and (ii) the ratios between the overlaps are related to the bulk-to-boundary OPE coefficients,
\begin{equation}
\label{eq:main_result2}
\frac{\bra{\psi^{a}_{\beta}} \phi^{\PBC}_{\alpha}\rangle}{\langle I^{a}| I^{\PBC}\rangle} = 2^{-2\Delta_\alpha+h_{\beta}} C^{a}_{\alpha\beta},
\end{equation}
where $C^{a}_{\alpha\beta}$ is the bulk-to-boundary OPE coefficient between the bulk operator $\phi_\alpha$ and the boundary operator $\psi^{a}_{\beta}$, $\Delta_{\alpha}$ is the scaling dimension of the bulk operator and $h_{\beta}$ is the scaling dimension of the boundary operator. Note also that Eqs.~\eqref{eq:main_result1} and \eqref{eq:main_result2} imply that all overlaps between the low-energy eigenstates $\bra{\psi^{a}_{\beta}} \phi^{\PBC}_{\alpha}\rangle$ decay as $N^{-c/16}$ at large sizes.

The rest of the paper is organized as follows: In Sec.~\ref{sect:CFT} we review the operator-state correspondence and bulk-to-boundary OPE. In Sec.~\ref{sect:overlap} we prove the main result Eqs.~\eqref{eq:main_result1} and \eqref{eq:main_result2} by writing the overlap as a path integral of the CFT. In Sec.~\ref{sect:examples} we numerically verify the main result by computing the overlaps for the Ising model and the three-state Potts model. Finally, we summarize the results and conclude with a few remarks on future directions in Sec.~\ref{sect:conclusion}. 

\section{Operator-state correspondence and bulk-to-boundary OPE}
\label{sect:CFT}
\subsection{Operator-state correspondence on the plane}
To begin with, we consider a CFT on the complex plane. The complex coordinates are $z=x+iy$ and $\bar{z}=x-iy$, where $x,y\in \mathbb{R}$. There is a set of primary operators $\phi_{\alpha}(z,\bar{z})$ which transform covariantly under conformal transformations $z\rightarrow z'=f(z)$, 
\begin{equation}
\label{eq:primary_op}
    \phi'_{\alpha}(z',\bar{z}')= \left(\frac{df}{dz}\right)^{-h_{\alpha}} \left(\frac{d\bar{f}}{d\bar{z}}\right)^{-\bar{h}_{\alpha}}\phi_{\alpha}(z,\bar{z}),
\end{equation}
where $h_{\alpha} (\bar{h}_{\alpha})$ is the (anti-)holomorphic conformal dimension. The scaling dimension is given by $\Delta_{\alpha} = h_{\alpha} + \bar{h}_{\alpha}$ and the conformal spin is given by $s_{\alpha} = h_{\alpha} - \bar{h}_{\alpha} \in \mathbb{Z}$. The operator-state correspondence implies that each primary operator $\phi_{\alpha}$ corresponds to a primary state $|\phi_{\alpha}\rangle$,
\begin{equation}
\label{eq:op_state_ket}
    |\phi_{\alpha}\rangle= \lim_{z,\bar{z}\rightarrow 0}\phi_{\alpha}(z,\bar{z})|I\rangle,
\end{equation}
where $|I\rangle$ is the vacuum state, defined as the unique state that is invariant under global conformal transformations. The conjugation of the ket state is
\begin{equation}
\label{eq:op_state_bra}
    \langle \phi_{\alpha}| = \lim_{z,\bar{z}\rightarrow\infty}\langle I| \phi_{\alpha}(z,\bar{z}) z^{2h_{\alpha}} \bar{z}^{2\bar{h}_{\alpha}}.
\end{equation}
The states form an orthonormal set,
\begin{equation}
    \langle \phi_{\alpha}|\phi_{\beta}\rangle = \delta_{\alpha\beta}.
\end{equation}
Next, we consider the CFT on the cylinder where the uncompactified direction $\tau\in(-\infty, \infty)$ represents imaginary time and the compactfied direction $X\in [0,L)$ represents space. The complex coordinates are $w=\tau+iX$ and $\bar{w}=\tau-iX$. We can use the conformal transformation
\begin{equation}
\label{eq:logmap}
    w = \frac{L}{2\pi} \log z
\end{equation}
to map the plane to the cylinder (abbreviated as $\cyl$ later on). The operator-state correspondence on the cylinder can be obtained by Eqs.~\eqref{eq:primary_op}-\eqref{eq:op_state_bra} and Eq.~\eqref{eq:logmap},
\begin{eqnarray}
\label{eq:op_state_cyl}
    |\phi^{\cyl}_{\alpha}\rangle &=& \phi^{\cyl}_{\alpha}(-\infty)|I^{\cyl}\rangle. \\
    \langle\phi^{\cyl}_{\alpha}| &=&  \langle I^{\cyl}| \phi^{\cyl}_{\alpha}(+\infty),
\end{eqnarray}
where
\begin{eqnarray}
    \phi^{\cyl}_{\alpha}(\pm \infty) \equiv \left(\frac{2\pi}{L}\right)^{-\Delta_{\alpha}} \lim_{\tau\rightarrow \pm\infty} e^{\pm \frac{2\pi}{L}\Delta_{\alpha}\tau} \phi^{\cyl}_{\alpha}(\tau,0).
\end{eqnarray}
This means that a primary state ket (bra) can be obtained by inserting the primary operator at the past (future) infinity. The state has energy and momentum \cite{cardy_conformal_1984,affleck_universal_1986,blote_1986}
\begin{eqnarray}
\label{eq:Ecyl}
    E^{\cyl}_{\alpha} &=& \frac{2\pi}{L} \left(\Delta_{\alpha}-\frac{c}{12}\right) \\
    P^{\cyl}_{\alpha} &=& \frac{2\pi}{L} s_{\alpha},
\end{eqnarray}
where $c$ is the central charge of the CFT. Note that the zero-point energy of the vacuum state is
\begin{equation}
\label{eq:CasmirPBC}
    E^{\cyl}_{I}(L)= -\frac{2\pi}{L} \frac{c}{12}.
\end{equation}
\subsection{Operator-state correspondence on the upper half plane}
The standard geometry for CFT with a boundary is the upper half plane with $x\in \mathbb{R}, y\in (0, \infty)$. The physics depends on the boundary condition, which we denote as $a$. Apart from the bulk operators $\phi_{\alpha}(z,\bar{z})$, there is a set of boundary primary operators $\psi^{a}_{\beta}(x)$ restricted on the boundary $y=0$. The set of boundary operators in general depends on the boundary condition $a$. Unlike bulk operators, the conformal spin of a boundary operator $\psi^{a}_{\beta}(x)$ cannot be defined, as rotation is not a symmetry of the upper half plane. Instead, a boundary operator is characterized by a single conformal dimension $h^{a}_{\beta}$, which is also known as the scaling dimension.  Under a conformal transformation $f$, a boundary primary operator transforms as
\begin{equation}
    \label{eq:primary_boundary_op}
      \psi'^{a}_{\beta}(f(x))= \left(\frac{df}{dx}\right)^{-h^{a}_{\beta}}\psi^{a}_{\beta}(x),
\end{equation}
The operator-state correspondence also works on the upper half plane,
\begin{eqnarray}
    |\psi^{a}_{\beta}\rangle &=& \lim_{x\rightarrow 0}\psi^{a}_{\beta}(x)|I^{a}\rangle.\\
    \langle \psi^{a}_{\beta}| &=& \lim_{x\rightarrow \infty}  x^{2h^{a}_{\beta}}\langle I^{a}|\psi^{a}_{\beta}(x),
\end{eqnarray}
where $|I^{a}\rangle$ is the vacuum state for the CFT with the boundary condition $a$. The states form an orthonormal set,
\begin{equation}
    \langle\psi^{a}_{\alpha}|\psi^{a}_{\beta}\rangle = \delta_{\alpha\beta}.
\end{equation}

One can use the conformal transformation
\begin{equation}
    w = \frac{L}{\pi} \log z
\end{equation}
to obtain a strip geometry (abbreviated as $\stp$ later on), where $w=\tau+iX$, $X\in (0,L)$ represents the space and $\tau\in (-\infty, \infty)$ represents the imaginary time. The boundaries are located at $X=0$ and $X=L$. Again, a state can be created by a boundary operator in the past (future) infinity of the strip,
\begin{eqnarray}
    |\psi^{a,\stp}_{\beta}\rangle &=& \psi^{a,\stp}_{\beta}(-\infty)|I^{a,\stp}\rangle. \\
    \label{eq:op_state_stp}
    \langle\psi^{a,\stp}_{\beta}| &=& \langle I^{a,\stp}| \psi^{a,\stp}_{\beta}(+\infty),
\end{eqnarray}
where $|I^{a,\stp}\rangle$ is the vacuum state on the strip and
\begin{equation}
    \psi^{a,\stp}_{\beta}(\pm \infty) \equiv \left(\frac{\pi}{L}\right)^{-h^{a}_{\beta}} \lim_{\tau\rightarrow \pm \infty} e^{\pm \frac{\pi}{L}\tau h^{a}_{\beta} }\psi^{a,\stp}_{\beta}(\tau,0)
\end{equation}
On the strip, the momenta cannot be defined, and the energies are related to the scaling dimensions by \cite{cardy_conformal_1984_2}
\begin{equation}
    E^{a,\stp}_{\beta} = \frac{\pi}{L}\left(h^{a}_{\beta}-\frac{c}{24}\right).
\end{equation}
In particular, the zero-point energy of the vacuum state is
\begin{equation}
\label{eq:CasimirOBC}
    E^{a,\stp}_{I}(L) = -\frac{\pi}{L} \frac{c}{24},
\end{equation}
which is independent of the boundary condition $a$. 

\subsection{Bulk-to-boundary OPE}
As is well known, the operator product expansion of bulk operators determines three and higher point correlation functions. In the presence of a boundary, there is a an additional type of OPE, the bulk-to-boundary OPE \cite{cardy_bulk_1991}. We focus on the geometry of the standard upper half plane, where the boundary is the real axis $y=0$. Given a bulk primary operator $\phi_{\alpha}(x,y)$, the bulk-to-boundary OPE is
\begin{equation}
    \phi_{\alpha}(x,y) = \sum_{\beta} (2y)^{-\Delta_{\alpha}+h_{\beta}} C^{a}_{\alpha\beta} \psi^{a}_{\beta}(x) + \cdots,
\end{equation}
where $C^{a}_{\alpha\beta}$ is the bulk-to-boundary OPE coefficient and $\cdots$ denotes contributions from descendant boundary operators. The bulk-to-boundary OPE plays a  crucial role in understanding boundary critical phenomena. For example, the boundary critical exponents are determined by the bulk-to-boundary OPE. The bulk-to-boundary OPE is also useful for evaluating the correlation functions consisting of a bulk operator and a boundary operator, e.g.
\begin{equation}
\label{eq:bbcorr}
    \langle \phi_{\alpha}(0,y) \psi^{a}_{\beta}(x)\rangle_{\UHP} = \frac{1}{|x|^{2h^{a}_{\beta}}} (2y)^{-\Delta_{\alpha}+h_{\beta}} C^{a}_{\alpha\beta}.
\end{equation}
In the limit of $x\rightarrow\infty$,
\begin{equation}
    \label{eq:bbcorr_infty}
        \langle \phi_{\alpha}(0,y) \psi^{a}_{\beta}(\infty)\rangle_{\UHP} =  (2y)^{-\Delta_{\alpha}+h_{\beta}} C^{a}_{\alpha\beta},
\end{equation}
where
\begin{equation}
    \psi^{a}_{\beta}(\infty) \equiv \lim_{x\rightarrow\infty} |x|^{2h^{a}_{\beta}} \psi^{a}_{\beta}(x).
\end{equation}
It is often interesting to focus on the bulk-to-boundary OPE where the boundary operator is the identity, $\psi^{a}_{\beta} = I$. Then Eq.~\eqref{eq:bbcorr} reduces to
\begin{equation}
     \langle \phi_{\alpha}(0,y) \rangle_{\UHP} = (2y)^{-\Delta_{\alpha}+h_{\beta}} C^{a}_{\alpha I},
\end{equation}
which is the expectation value of a bulk primary operator in the presence of a boundary. By convention, the bulk-to-boundary OPE involving the identity operator is denoted by $A^{a}_{\alpha} \equiv C^{a}_{\alpha I}$.
\subsection{Boundary conditions for rational CFTs}
For rational conformal field theories, where the number of primary operators is finite, a complete classification of boundary conditions together with all possible boundary operators was found by Cardy \cite{cardy_boundary_1989}. For simplicity we only consider diagonal CFTs, where each bulk primary operator has $h_{\alpha}=\bar{h}_{\alpha}$. The result is briefly summarized as follows. A physical boundary condition is labelled by a bulk primary operator. The label $a$ for the boundary condition thus runs over bulk primary operators. All boundary primary operators correspond to the chiral component of a bulk primary operator. Thus the label $\beta$ in the boundary operator $\psi^{a}_{\beta}$ also runs over bulk primary operators. A boundary operator can appear on the boundary with boundary condition$a$ only if the corresponding bulk operator $\phi_{\beta}$ is in the bulk-to-bulk OPE $\phi_{a}\times \phi^{\dagger}_{a}$, where $\phi_{a}$ is the bulk primary operator corresponding to the boundary condition $a$ and $\phi^{\dagger}_a$ is its conjugate.

The bulk-to-boundary OPE can be obtained from the Moore-Seiberg data of a rational CFT. In particular, the bulk-to-boundary OPE coefficient $A^{a}_{\alpha} \equiv C^{a}_{\alpha I}$ involving the identity boundary operator can be computed from
\begin{equation}
\label{eq:A_Smatrix}
    A^{a}_{\alpha}=\frac{S_{\alpha a}}{S_{Ia}}\sqrt{\frac{S_{II}}{S_{\alpha I}}},
\end{equation}
where $S$ is the modular $S$ matrix in the Moore-Seiberg data. General bulk-to-boundary OPE cofficients can also be computed from the Moore-Seiberg data but in a more complicated way \cite{lewellen_sewing_1992}. 

\section{Path integral for overlaps}
\label{sect:overlap}
In this section, we derive the main result Eqs.~\eqref{eq:main_result1} and \eqref{eq:main_result2}. We consider a quantum critical spin chain whose continuum limit can be described by a CFT. In practice, it is observed that physical observables on the lattice differ from their continuum counterpart by a finite-size correction that decreases with the system size $N$. We will therefore proceed by considering the continuum counterpart of the overlaps and assume that analogous results hold for the quantum spin chain with sufficiently large sizes.

The strategy to derive the result is as follows. First we express the overlaps as a path integral of the CFT on a Riemann surface. Then we use a conformal mapping to map it to the standard geometries, i.e., the strip, the cylinder and the upper half plane discussed in the last section. Finally, we express the path integrals on the standard geometries as a function of the conformal data, which leads to the main result Eqs.~\eqref{eq:main_result1} and \eqref{eq:main_result2}.
\subsection{Overlap between ground states}
\begin{figure}
    \centering
    \includegraphics[width=1.0\linewidth]{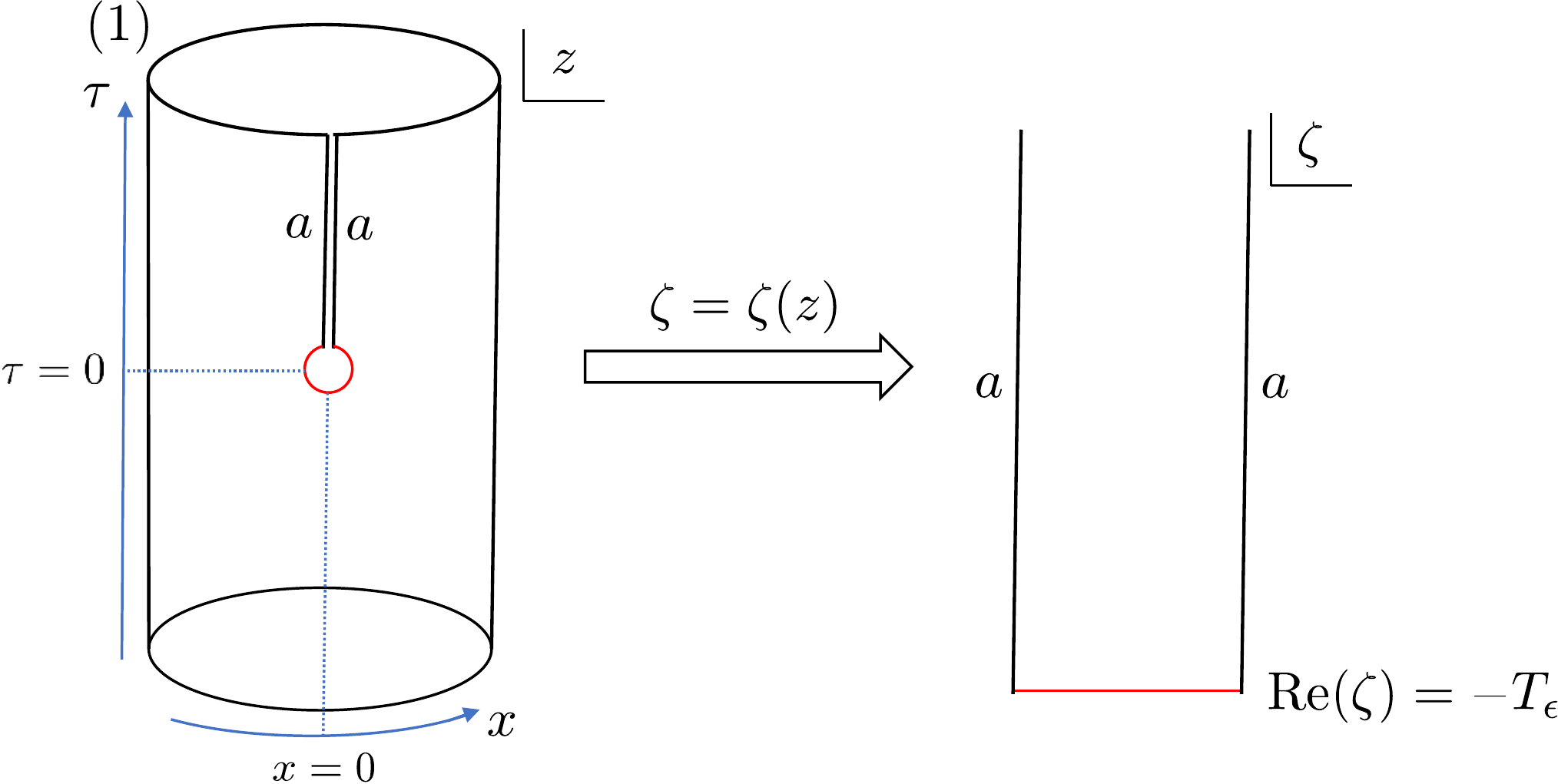}
    \includegraphics[width=1.0\linewidth]{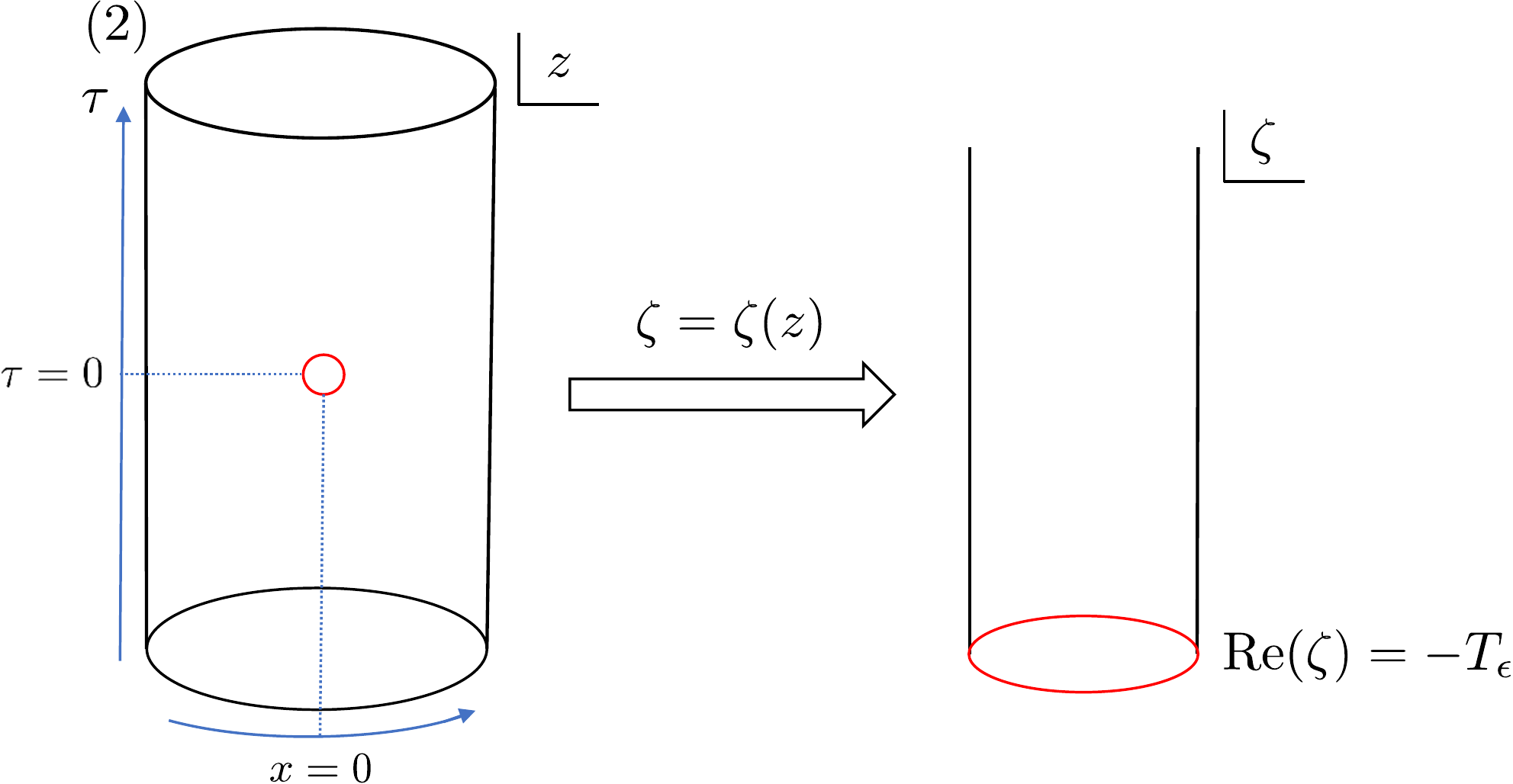}
    \includegraphics[width=1.0\linewidth]{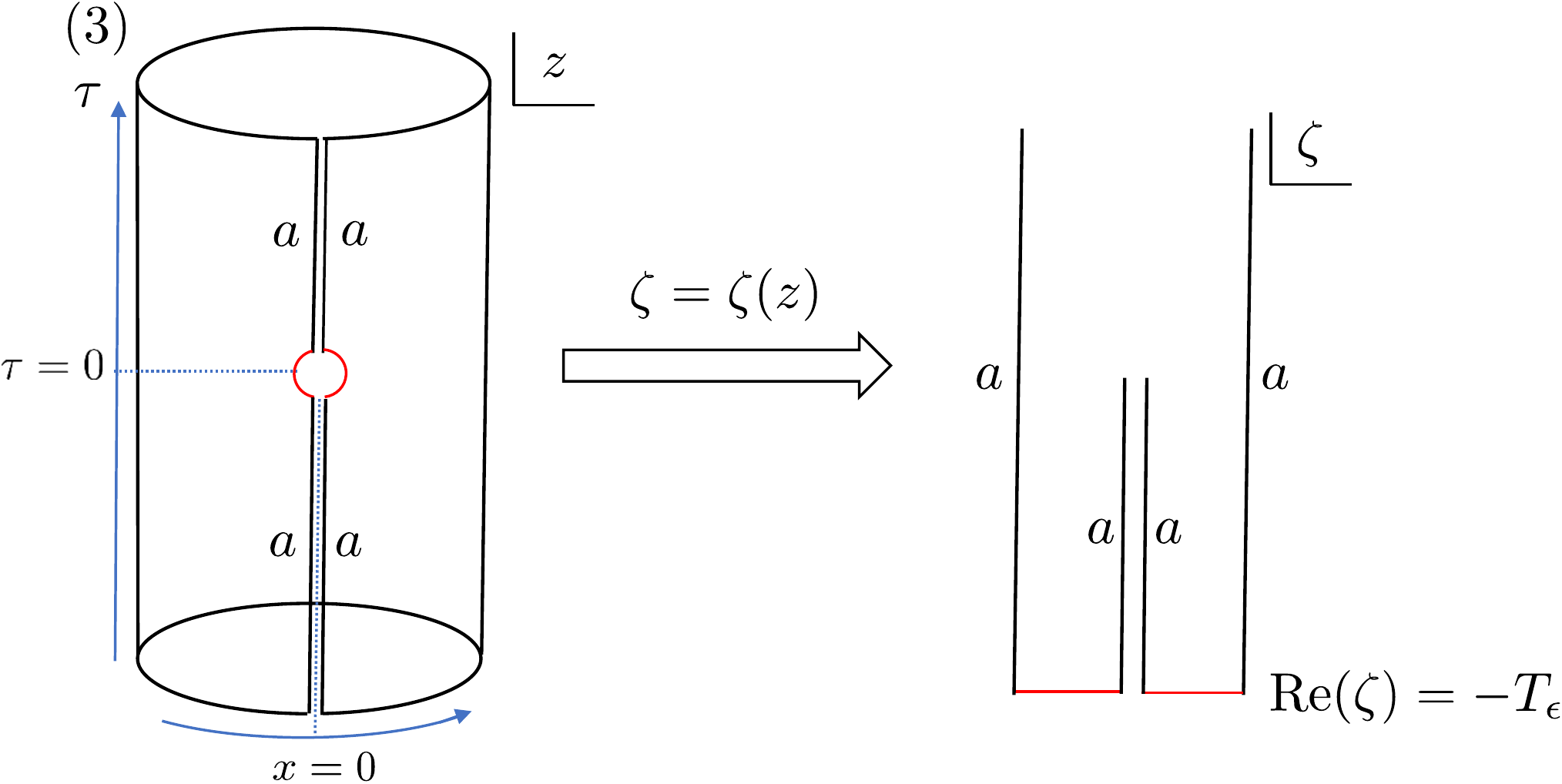}
    \caption{Geometries for the path integrals in Eq.~\eqref{eq:identity_overlaps} is presented on the left. The conformal mapping Eq.~\eqref{eq:conformal_map1} is used to map the geometries into strip-like geometries.}
    \label{fig:Zs}
\end{figure}

To begin with, consider a cylinder of circumference $L$. The Hilbert space is supported on equal-time slices. On the time slice $\tau=0$, the Hilbert space is spanned by the basis $\{|\phi(x)\rangle\}$, where $\phi(x)$ is the value of the fundamental field $\Phi(\tau,x)$ on the $\tau=0$ surface. The ground state $|I^{\cyl}\rangle$ is obtained by the path integral on the $\tau<0$ half of the cylinder, such that the wavefunctional reads
\begin{equation}
    \langle \phi(x)|I^{\cyl}\rangle = \frac{1}{\sqrt{Z^{\cyl}}}\int_{\substack{\tau<0, ~cyl \\ \Phi(0,x) = \phi(x)}} D\Phi\, e^{-S},
\end{equation}
where $S$ is the action of the CFT, $D\Phi$ is the functional measure, and
\begin{equation}
    Z^{\cyl} = \int_{cyl} D\Phi\, e^{-S}
\end{equation}
is the partition function. Analogously, the conjuagte of the the ground state $\langle I^{a,\stp}|$ in the strip geometry is also obtained by a path integral on the $\tau>0$ half of the strip, such that the wavefunctional reads
\begin{equation}
        \langle I^{a,\stp}|\phi(x)\rangle = \frac{1}{\sqrt{Z^{a}}}\int_{\substack{\tau>0, ~stp \\ \Phi(0,x) = \phi(x)}} D\Phi\, e^{-S},
\end{equation}
where $Z^{a}$ is the partition function on a strip with width $L$ and boundary condition $a$.
The overlap is obtained by gluing the two geometries through the $\tau=0$ interface,
\begin{equation}
\label{eq:Idoverlap_noUVcutoff}
    \langle I^{a,\stp}|I^{\cyl}\rangle \sim \frac{1}{\sqrt{Z^{a}Z^{\cyl}}}\int_{\substack{\tau>0, ~stp \\ \tau<0, ~cyl}} D\Phi\, e^{-S}.
\end{equation}
However, the gluing results in a singular geometry where the boundary condition changes sharply across the $\tau=0$ interface. In order to avoid the singularity, we instead consider the geometry labelled as (1) in Fig.~\ref{fig:Zs}, where a circle with a small radius $\epsilon$ centered at $\tau=x=0$ is removed from the path integral. The radius $\epsilon$ represents the UV cutoff, analogous to the UV cutoff induced by an entanglement cut in a CFT \cite{calabrese_entanglement_2004,Cardy_entanglement_2016}. The precise boundary condition on the circle does not play a role in the computation below. Such a UV cutoff corresponds to a finite lattice spacing in the spin chain. The retio between the circumference $L$ of the cylinder (or the width of the strip) and the UV cutoff $\epsilon$ corrsponds to the number of spins $N$ in a spin chain,
\begin{equation}
    N \sim L/\epsilon.
\end{equation}
The thermodynamic limit $N\rightarrow \infty$ corresponds to the limit $\epsilon\rightarrow 0$ in the path integral. The same UV cutoff is introduced in the partition function for the CFT on the cylinder and on the strip, which is shown in subfigures (2) and (3) in Fig.~\ref{fig:Zs}. Let $i=1,2,3$ label the three geometries (1), (2), (3) in Fig.~\ref{fig:Zs}, and the partition function on the three geometries be
\begin{equation}
    Z_i=\int_{i} D\phi\, e^{-S},
\end{equation}
then the UV regularized version of the overlap Eq.~\eqref{eq:Idoverlap_noUVcutoff} is
\begin{equation}
\label{eq:identity_overlaps}
    \bra{I^{a}} I^{\PBC}\rangle=\frac{Z_1}{\sqrt{Z_2Z_3}},
\end{equation}
where the denominator accounts for the normalization of the states. Next, we use a conformal mapping to transform the geometries (1)-(3) into strip-like geometries. This is acheived by
\begin{eqnarray}
\label{eq:conformal_map1}
    \zeta(z)=\log (e^{2\pi z/L}-1),
\end{eqnarray}
where the resulting geometries are shown in Fig.~\ref{fig:Zs}. Note that the partition function is not invariant under conformal transformations, since the metric $g$ transforms as $g_{ab}\rightarrow e^{2\phi} g_{ab}$. The partition function takes on an additional factor $Z\rightarrow e^{-S_L[\phi,g]} Z$ where $S_L[\phi,g]$ is the Liouville action \cite{caputa_liouville_2017,polyakov_quantum_1981}. However, the ratio Eq.~\eqref{eq:identity_overlaps} is invariant under conformal transformations, since the Liouville factor cancels out. 

Under the conformal mapping Eq.~\eqref{eq:conformal_map1}, all three geometries are transformed into a strip-like geometry with width $2\pi$. The circle representing the UV cutoff is mapped onto a horizontal cutoff at $\mathrm{Re} ~\zeta = -T_{\epsilon} = \log (2\pi \epsilon/L)$. As the system size changes, $N\rightarrow bN$,  $T_{\epsilon}$ goes to $T_{\epsilon}+ \log b$. In the limit of $T_{\epsilon}\rightarrow\infty$, the change in the partition functions is dominated by the ground state in the corresponding strip-like geometries,
\begin{eqnarray}
    Z_i \rightarrow Z_i e^{-E_i \log b} 
\end{eqnarray}
where $E_1 = E^{a,\stp}_I(2\pi)$ is the Casimir energy of a CFT on a strip with width $2\pi$, $E_2 = E^{\cyl}_I(2\pi)$ is the Casimir energy on a cylinder with circumference $2\pi$, and $E_3=2 E^{a,\stp}_I(\pi)$ is the Casimir energy on two strips with width $\pi$. Using Eqs.~\eqref{eq:CasmirPBC} and \eqref{eq:CasimirOBC}, we obtain
\begin{eqnarray}
    E_1 &=& -\frac{c}{48} \\
    E_2 &=& -\frac{c}{12} \\
    E_3 &=& -\frac{c}{12}.
\end{eqnarray}
Finally, the overlap between ground states transforms as
\begin{equation}
\label{eq:overlapRG}
    \frac{Z_1}{\sqrt{Z_2Z_3}}\rightarrow b^{-[E_1- (E_2+E_3)/2]}\frac{Z_1}{\sqrt{Z_2Z_3}},
\end{equation}
under the rescaling $N\rightarrow bN$.
The overlap between ground states thus scales as
\begin{equation}
    \bra{I^{a}} I^{\PBC}\rangle \propto N^{-[E_1- (E_2+E_3)/2]}.
\end{equation}
Therefore, at large system sizes
\begin{equation}
    \bra{I^{a}} I^{\PBC}\rangle = \mathcal{N}_a N^{-c/16}.
\end{equation}
\subsection{Overlap between other primary states}
\begin{figure}
    \centering
    \includegraphics[width=1.0\linewidth]{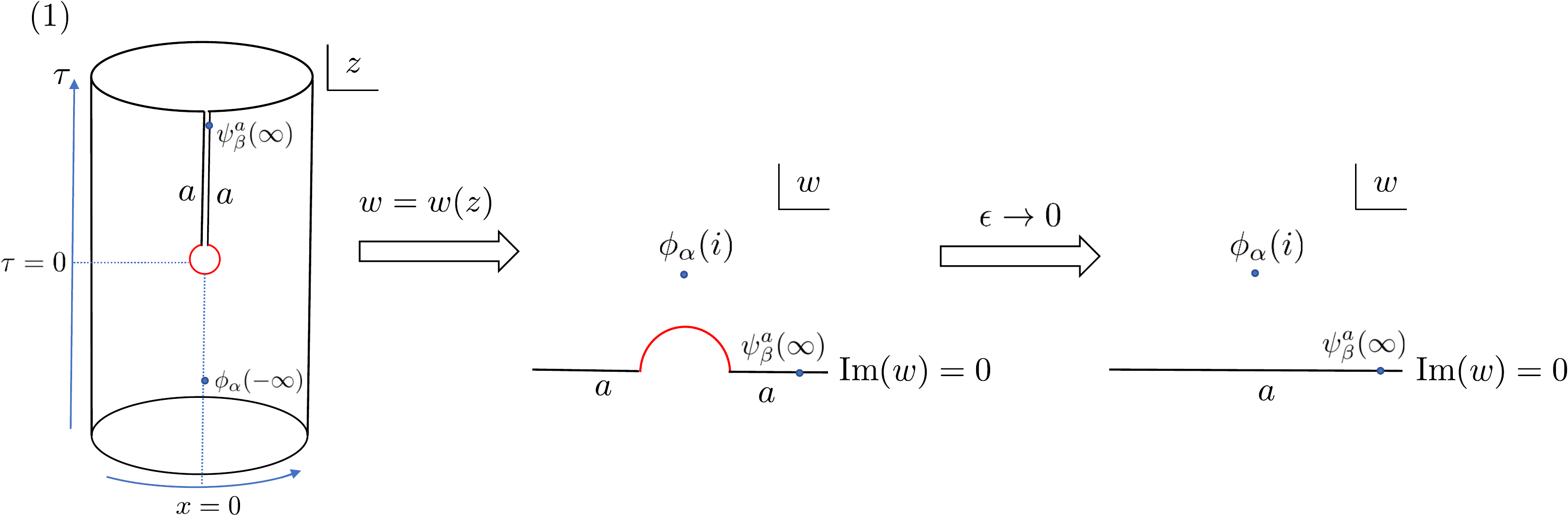}
    \caption{The path integral for the overlap Eq.~\eqref{eq:ab_overlaps} is shown on the left. The conformal mapping Eq.~\eqref{eq:conformal_mapping_2} is used to map the geometry into the middle geometry. Taking the thermodynamic limit $\epsilon\rightarrow 0$ we obtain the upper half plane (UHP) on the right.}
    \label{fig:2ptcorrs}
\end{figure}

The operator-state correspondence can be expressed in terms of path integrals. On the cylinder, Eq.~\eqref{eq:op_state_cyl} means that a primary state $|\phi^{cyl}_{\alpha}\rangle$ can be prepared by inserting a primary operator in the path integral at the past infinity, such that the wavefunctional reads
\begin{eqnarray}
    &~&\langle \phi(x)|\phi^{cyl}_{\alpha}\rangle   \nonumber \\
    &=&  \frac{1}{\sqrt{Z^{cyl}}} \int_{\substack{\tau<0, ~cyl \\ \Phi(0,x) = \phi(x)}} D\Phi\, \phi^{\cyl}_{\alpha}(-\infty) e^{-S}.
\end{eqnarray}
Analogously, the wavefunctional of a primary state $\langle \psi^{a,stp}_{\beta}|$ on the strip can be represented as a path integral,
\begin{eqnarray}
        &~&\langle \psi^{a,stp}_{\beta}|\phi(x)\rangle \nonumber \\
        &=&\frac{1}{\sqrt{Z^{a}}} \int_{\substack{\tau>0, ~stp \\ \Phi(0,x) = \phi(x)}} D\Phi\, \psi^{a,stp}_{\beta}(+\infty) e^{-S}.
\end{eqnarray}
In order to compute the overlap, we glue the two path integrals through the $\tau=0$ interface, which gives
\begin{eqnarray}
    &~&\langle \psi^{a,stp}_{\beta} |\phi^{\cyl}_{\alpha}\rangle \\
    \label{eq:overlapab} &=& \frac{1}{\sqrt{Z^{\cyl}Z^{a}}} 
      \int_{\substack{\tau<0, ~cyl \\ \tau>0, ~stp}} D\Phi \phi^{\cyl}_{\alpha}(-\infty) \psi^{a,\stp}_{\beta}(+\infty)e^{-S}.
\end{eqnarray}
Again, the path integral is on a singular geometry, and we remove a circle with radius $\epsilon$ representing the UV cutoff, which gives the geometry (1) in Fig.~\ref{fig:2ptcorrs}. 
Then the UV regularized version of the overlap is
\begin{equation}
\label{eq:ab_overlaps}
    \langle \psi^{a}_{\beta} |\phi^{\PBC}_{\alpha}\rangle = \frac{Z_1}{\sqrt{Z_2Z_3}} \langle \phi_{\alpha}(-\infty)\psi^{a}_{\beta}(+\infty) \rangle_{1},
\end{equation}
where we have used the fact that the correlation function of a bulk operator and a boundary operator on the geometry (1) is
\begin{equation}
     \langle \phi_{\alpha}(-\infty)\psi^{a}_{\beta}(+\infty) \rangle_{1} = \frac{1}{Z_1}\int_{1} D\Phi\, \phi_{\alpha}(-\infty)\psi^{a}_{\beta}(+\infty) e^{-S}.
\end{equation}
Combining Eq.~\eqref{eq:ab_overlaps} and Eq.~\eqref{eq:identity_overlaps}, we obtain the ratio between overlaps,
\begin{equation}
\label{eq:overlap_ratio}
    \frac{\langle \psi^{a}_{\beta} |\phi^{\PBC}_{\alpha}\rangle}{\langle I^{a} |I^{\PBC}\rangle} = \langle \phi_{\alpha}(-\infty)\psi^{a}_{\beta}(+\infty) \rangle_{1}.
\end{equation}
In the limit of $\epsilon\rightarrow 0$, the conformal mapping
\begin{equation}
\label{eq:conformal_mapping_2}
    w=\sqrt{e^{2\pi z/L}-1}
\end{equation}
transforms the geometry into the upper half plane. The fact that we are working with finite $\epsilon$ only contributes to the finite-size correction, as examined in the appendix. The bulk and boundary operators transform according to Eqs.~\eqref{eq:primary_op} and \eqref{eq:primary_boundary_op}. Therefore,
\begin{eqnarray}
    &~& \langle \phi_{\alpha}(-\infty)\psi^{a}_{\beta}(+\infty) \rangle_{1} \nonumber \\ \label{eq:standard_bb_corr_trans}
    &=& |w'(-\infty)|^{\Delta_{\alpha}}(w'(+\infty))^{h_{\beta}}\langle \phi_{\alpha}(i) \psi_{\beta}(\infty)\rangle_{\UHP},
\end{eqnarray}
where
\begin{eqnarray}
    w'(-\infty) &=& \left(\frac{2\pi}{L}\right)^{-1}\lim_{\tau\rightarrow -\infty} w'(\tau)e^{-\frac{2\pi}{L}\tau}\\
    w'(+\infty) &=& \left(\frac{\pi}{L}\right)^{-1}\lim_{\tau\rightarrow +\infty} w'(\tau)e^{-\frac{\pi}{L}\tau}.
\end{eqnarray}
Working out the limit, we obtain
\begin{eqnarray}
\label{eq:diff_minf}
    w'(-\infty) &=& \frac{1}{2i} \\
    \label{eq:diff_pinf}
    w'(+\infty) &=& 1
\end{eqnarray}
Finally, the correlation function on the upper half plane is given by Eq.~\eqref{eq:bbcorr_infty}, 
\begin{eqnarray}
\label{eq:standard_bbcorr_uhp}
    \langle\phi_{\alpha}(i) \psi^{a}_{\beta}(\infty)\rangle_{\UHP}= 2^{-\Delta_{\alpha}+h_{\beta}} C^{a}_{\alpha\beta}.
\end{eqnarray}
Substituting Eqs.~\eqref{eq:diff_minf}-\eqref{eq:standard_bbcorr_uhp} and Eq.~\eqref{eq:standard_bb_corr_trans} into Eq.~\eqref{eq:overlap_ratio} we obtain
\begin{equation}
\frac{\bra{\psi^{a}_{\beta}} \phi^{\PBC}_{\alpha}\rangle}{\langle I^{a}| I^{\PBC}\rangle} = 2^{-2\Delta_\alpha+h_{\beta}} C^{a}_{\alpha\beta}.
\end{equation}

\section{Examples}
\label{sect:examples}
In this section we test the main result in two lattice models, the Ising model and the three-state Potts model.
\subsection{Setup}
For simplicity, we restrict our attention to critical quantum spin chains with nearest-neighbor interactions. The Hamiltonian density is composed of a two-site term $h^{(2)}$ and a one-site term $h^{(1)}$. For PBC, the Hamiltonian is
\begin{equation}
    H^{\PBC}=\sum_{j=1}^{N} h^{(2)}_{j,j+1}+\sum_{j=1}^{N}h^{(1)}_{j},
\end{equation}
where site $j=N+1$ is identified with site $j=1$. The low-energy eigenstates are in one-to-one correspondence with the scaling operators in the CFT. The energies and momenta at low energies are similar to those of the CFT on the cylinder,
\begin{eqnarray}
    E^{\PBC}_{\alpha} &=& A+\frac{B}{N} \left(\Delta_{\alpha}-\frac{c}{12}\right)+O\left(\frac{1}{N^x}\right) \\
    P^{\PBC}_{\alpha} &=& \frac{2\pi}{N} s_{\alpha},    
\end{eqnarray}
where $A,B$ and $x>1$ are non-universal constants. The primary states $|\phi^{\PBC}_{\alpha}\rangle$ can be identified using the method in Ref.~\onlinecite{milsted_extraction_2017,zou_conformal_2018}. 

Next, we consider the same model on an open chain with the same boundary condition $a$ on both ends. The Hamiltonian is
\begin{equation}
    H^{a} = \sum_{j=1}^{N-1} h^{(2)}_{j,j+1}+\sum_{j=2}^{N-1}h^{(1)}_{j}+h^{a}_L+h^{a}_R,
\end{equation}
where $h^{a}_{L/R}$ is localized near site $1$ or site $N$. The low-energy spectrum is similar to that of the CFT on the strip,
\begin{equation}
    E^{a}_{\beta} = A^{a} +\frac{B}{2N} \left(h^{a}_{\beta}-\frac{c}{24}\right)+O\left(\frac{1}{N^x}\right),
\end{equation}
where $A^{a}$ is a non-universal constant depending on the boundary condition. The low-energy states are in on-to-one correspondence with a boundary operator, and we may identify the primary states $|\psi^{a}_{\beta}\rangle$ that correspond to boundary primary operators.

Different choices of the boundary term $h^{a}$ may flow to the same conformal boundary condition in the CFT. Moreover, the identification of a lattice boundary condition with a CFT boundary condition is not known a priori. In practice, one may compute the low-energy spectrum $E^{a}_{\beta}$ at sufficiently large sizes to read the spectrum of boundary operators, which is determined by the conformal boundary condition.

In the following we focus on the Ising model and the three-state Potts model. In each subsection, we first review relevant CFT data and then present numerical results for the overlaps. For each lattice model, we first diagonalize the low-energy eigenstates of the model with PBC and open ends. In order to go to large sizes, we use matrix product state methods \cite{white_1992,rommer_1997,zou_conformal_2018} (see also appendix for more details). We will compute the overlap $\langle I^{a}|I^{\PBC}\rangle$ for a series of system sizes and extract the exponent in the right-hand side of Eq.~\eqref{eq:main_result1}. Then we compute other overlaps $\langle \psi^{a}_{\beta}|\phi^{\PBC}_{\alpha}\rangle$ divided by the ground-state overlap. The ratios are then extrapolated to the thermodynamic limit, where we compare with the right-hand side of Eq.~\eqref{eq:main_result2}. The extrapolation is performed by fitting the finite-size data with the formula $y=p_1+p_2 N^{-p_3}$, where $p_1,p_2$ and $p_3>0$ are constants to be fixed.

There is a phase ambiguity of the eigenstates which can be fixed in the following way. The overlap $\langle I^{a}|I^{\PBC}\rangle$ is taken to be real and positive for all boundary conditions $a$. For an eigenstate $|\phi^{\PBC}_{\alpha}\rangle$ of the spin chain with PBC, we require $\langle I^{I}|\phi^{\PBC}_{\alpha}\rangle$ to be real and positive. This is in accordance with Eq.~\eqref{eq:main_result2} and the fact that $S_{I\alpha}>0$ for unitary CFTs. Finally, for other eigenstates of the spin chain with open ends, we choose to fix the phase by requiring one of the nonzero overlaps $\langle \psi^{a}_{\beta}|\phi^{\PBC}_{\alpha}\rangle$ to be real and positive. The choice can be arbitrary and correspond to a specific choice of the normalization of the boundary operator $\psi^{a}_{\beta}$.

\subsection{Ising model}
The Ising model is described by a unitary minimal CFT with central charge $c=1/2$ \cite{belavin_infinite_1984}. There are three chiral primary fields in the Ising CFT, as shown in Table \ref{tab:Ising_chiral}.
\begin{table}[htbp]
    \centering
    \begin{tabular}{|c|c|}
    \hline
         Chiral field & $h$    \\ \hline
           $I$ &  $0$    \\ \hline
           $\sigma$ & $1/16$ \\ \hline
           $\psi$ & $1/2$ \\ \hline
    \end{tabular}
    \caption{Chiral fields in Ising CFT}
    \label{tab:Ising_chiral}
\end{table}
The boundary primary operators are in one-to-one correspondence with the chiral primary fields. We will also denote the corresponding boundary operators by $I, \sigma$ and $\psi$. The bulk primary operators are diagonal combinations of chiral and anti-chiral primary fields, as shown in Table \ref{tab:Ising_bulk}.
\begin{table}[htbp]
    \centering
    \begin{tabular}{|c|c|c|c|}
    \hline
         Bulk operator & $\Delta=h+\bar{h}$ & $s=h-\bar{h}$ & $\mathbb{Z}_2$ charge   \\ \hline
           $I$ &  $0$  & $0$ & 0\\ \hline
           $\sigma$ & $1/8$ & $0$ & $1$ \\ \hline
           $\varepsilon$ & $1$ & $0$ & $0$\\ \hline
    \end{tabular}
    \caption{Bulk primary operators in the Ising CFT}
    \label{tab:Ising_bulk}
\end{table}

Conformal boundary conditions are in one to one correspondence with the bulk primary operators \cite{cardy_effect_1986,cardy_boundary_1989}. We will denote the boundary conditions as $a=I,\sigma,\varepsilon$. The spectrum of boundary operators is determined by the boundary condition. For both $I$ and $\varepsilon$ boundary conditions, there is only one boundary primary operator $I$. For the $\sigma$ boundary condition, there are two boundary primary operators, $I$ and $\psi$.

In order to compute the bulk-to-boundary OPE coefficients, we need the modular $S$ matrix in the Moore-Seiberg data \cite{moore_classical_1989}. For the Ising CFT we have
\begin{equation}
\label{eq:Smatrix_Ising}
    S=\begin{bmatrix}
       1/2 & 1/\sqrt{2} & 1/2 \\
       1/\sqrt{2} & 0 & -1/\sqrt{2} \\
       1/2 & -1/\sqrt{2} & 1/2
\end{bmatrix}
\end{equation}
Substituting Eq.~\eqref{eq:Smatrix_Ising} into Eq.~\eqref{eq:A_Smatrix} we obtain the bulk-to-boundary OPE coefficients,
\begin{eqnarray}
\label{eq:Isingbb1}
    A^{I}_{\sigma}=-A^{\varepsilon}_{\sigma} = 2^{1/4}, ~~ A^{I}_{\varepsilon}= A^{\varepsilon}_{\varepsilon}=1 \\
    A^{\sigma}_{\sigma}=0, ~~ A^{\sigma}_{\varepsilon}=-1.
\end{eqnarray}
Note also that $A^{a}_I=1$ for all boundary conditions. 

The CFT has a $\mathbb{Z}_2$ symmetry. In the bulk, the $I$ and $\varepsilon$ fields are even under $\mathbb{Z}_2$ symmetry while the $\sigma$ field is odd. The boundary condition $\sigma$ is even under $\mathbb{Z}_2$ and the $I$ and $\varepsilon$ boundary conditions transform into each other under the symmetry. The symmetry is manifest in the bulk-to-boundary OPE coefficients in Eq.~\eqref{eq:Isingbb1}.

There is one nonzero bulk-to-boundary OPE that involves the boundary operator $\psi$ in the $\sigma$ boundary condition \cite{cardy_bulk_1991},
\begin{equation}
    C^{\sigma}_{\sigma\psi} = +2^{-1/4},
\end{equation}
where the $+$ sign reflects a specific choice of the normalization of the boundary operator $\psi^{\sigma}$. Note that $C^{\sigma}_{\sigma\psi}$ is not to be confused with the bulk-to-bulk or boundary-to-boundary OPE coefficients, as the upper index labels the boundary condition rather than a field.

Now we turn to the lattice model. Let $X_j ~(Z_j)$ be the Pauli $X ~(Z)$ operator on site $j$. The Hamiltonian density of the Ising model is 
\begin{equation}
    h^{(2)}_{j,j+1}=-X_j X_{j+1}, ~~ h^{(1)}_j = -Z_j.
\end{equation}
One lattice realization of the conformal boundary conditions is given by the free and fixed boundary conditions \cite{cardy_effect_1986},
\begin{eqnarray}
    h^{I}_{L/R} &=& \beta X_{1/N},~~~\, \mathrm{(Fixed~up)}\\
    h^{\sigma}_{L/R} &=& 0, ~~~~~~~~~~~\mathrm{(Free)} \\
    h^{\varepsilon}_{L/R} &=& -\beta X_{1/N}, ~\mathrm{(Fixed~down)}
\end{eqnarray}
where $\beta>0$ is arbitrary. In the numerical simulation it is convenient to choose $\beta$ on order $1$, as a too small $\beta$ renders large finite-size corrections and too large $\beta$ may induce numerical instability. We will use $\beta=4$ in the numerical simulations below. 

The Ising model has a $\mathbb{Z}_2$ symmetry which transforms $X_i\rightarrow -X_i, Z_i\rightarrow Z_i$. The free boundary condition preserves the $\mathbb{Z}_2$ symmetry. The fixed boundary conditions are transformed into each other by the symmetry. We will therefore focus on the free and fixed-up boundary conditions. 

For Ising model with PBC, one can identify the ground state as $|I^{\PBC}\rangle$, the first excited state as $|\sigma^{\PBC}\rangle$ and the second excited state as $|\varepsilon^{\PBC}\rangle$. The states $|I^{\PBC}\rangle$ and $|\varepsilon^{\PBC}\rangle$ are even under $\mathbb{Z}_2$ while $|\sigma^{\PBC}\rangle$ is odd. In the fixed-up boundary conditions, one can identify the ground state as $|I^{I}\rangle$. In the free boundary condition, the ground state is identified with $|I^{\sigma}\rangle$ and the first excited state is identified as $|\psi^{\sigma}\rangle$. 

\begin{figure}[hthp]
    \centering
    \includegraphics[width=0.98\linewidth]{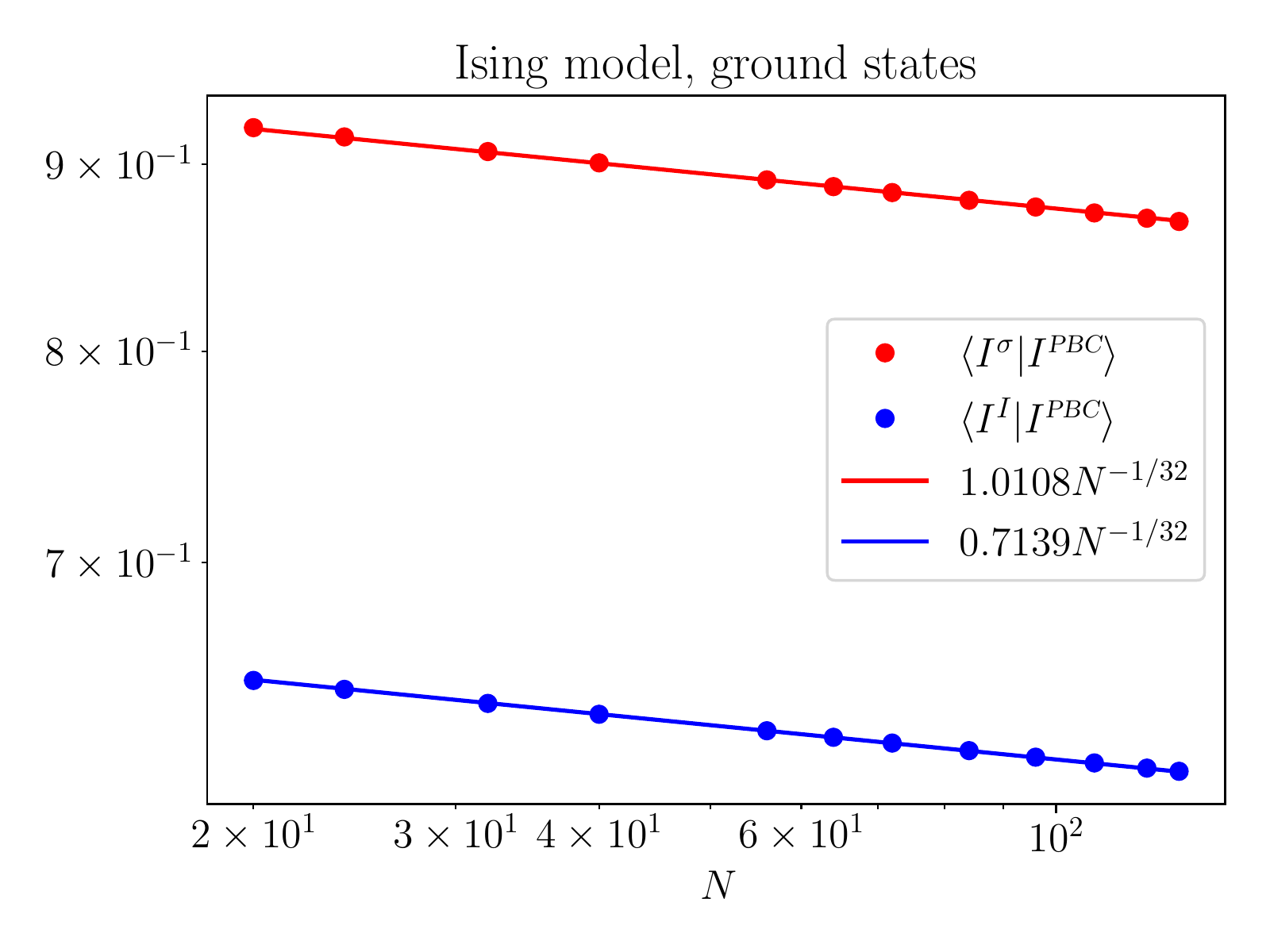}
    \caption{Overlaps between ground states of the Ising model with periodic boundary conditions and open ends. The solid lines represent exponential decaying functions with the correct exponent $-c/16=-1/32$. The data is obtained by numerical simulations of the Ising model with $20 \leq N \leq 128$.}
    \label{fig:Ising1}
\end{figure}

For the overlap between ground states, we substitute $c=1/2$ into Eqs.~\eqref{eq:main_result1} to obtain
\begin{eqnarray}
\label{eq:main_result1_Ising}
    \bra{I^{a}} I^{\PBC}\rangle &=& \mathcal{N}_{a} N^{-1/32}, ~a=I,\sigma.
\end{eqnarray}
Overlaps that are related to the bulk-to-boundary OPEs are shown in Table.~\ref{tab:Ising_OPE}. 
\begin{table}[htbp]
    \begin{tabular}{|c|c|c|c|}
    \hline
           $\phi^{\PBC}_{\alpha}$ & $\psi^{a}_{\beta}$ & \makecell{$\bra{\psi^{a}_{\beta}} \phi^{\PBC}_{\alpha}\rangle/\langle I^{a}| I^{\PBC}\rangle$ \\ (Numerical value)} & \makecell{$2^{-2\Delta_\alpha+h_{\beta}} C^{a}_{\alpha\beta}$ \\ (CFT data)} \\ \hline
           $\sigma$ & $I^{I}$ & 0.999&  1 \\ \hline
           $\sigma$ & $\psi^{\sigma}$ & 0.999& 1\\ \hline
           $\sigma$ & $I^{\sigma}$ & 0 & 0 \\ \hline
           $\varepsilon$ & $I^{I}$ &0.247 & 1/4\\ \hline
           $\varepsilon$ & $I^{\sigma}$ &$-0.251$ & $-1/4$ \\ \hline
    \end{tabular}
    \caption{The ratios of the overlaps for the Ising model. The first two columns specify the two states in the overlap, and the third column shows the ratio $\bra{\psi^{a}_{\beta}} \phi^{\PBC}_{\alpha}\rangle/\langle I^{a}| I^{\PBC}\rangle$ extrapolated to the thermodynamic limit (LHS of Eq.~\eqref{eq:main_result2}). The fourth column shows the corresponding CFT quantity $2^{-2\Delta_\alpha+h_{\beta}} C^{a}_{\alpha\beta}$ (RHS of Eq.~\eqref{eq:main_result2}).}
    \label{tab:Ising_OPE}
\end{table}
The overlaps between ground states are shown in Fig.~\ref{fig:Ising1}. We see that both overlaps decay as $N^{-1/32}$ as expected. The coefficients $\mathcal{N}_{I}$ and $\mathcal{N}_{\sigma}$ are non-universal. They depend on specific lattice realizations of the CFT and the conformal boundary condition.

\begin{figure}[hthp]
    \centering
    \includegraphics[width=0.98\linewidth]{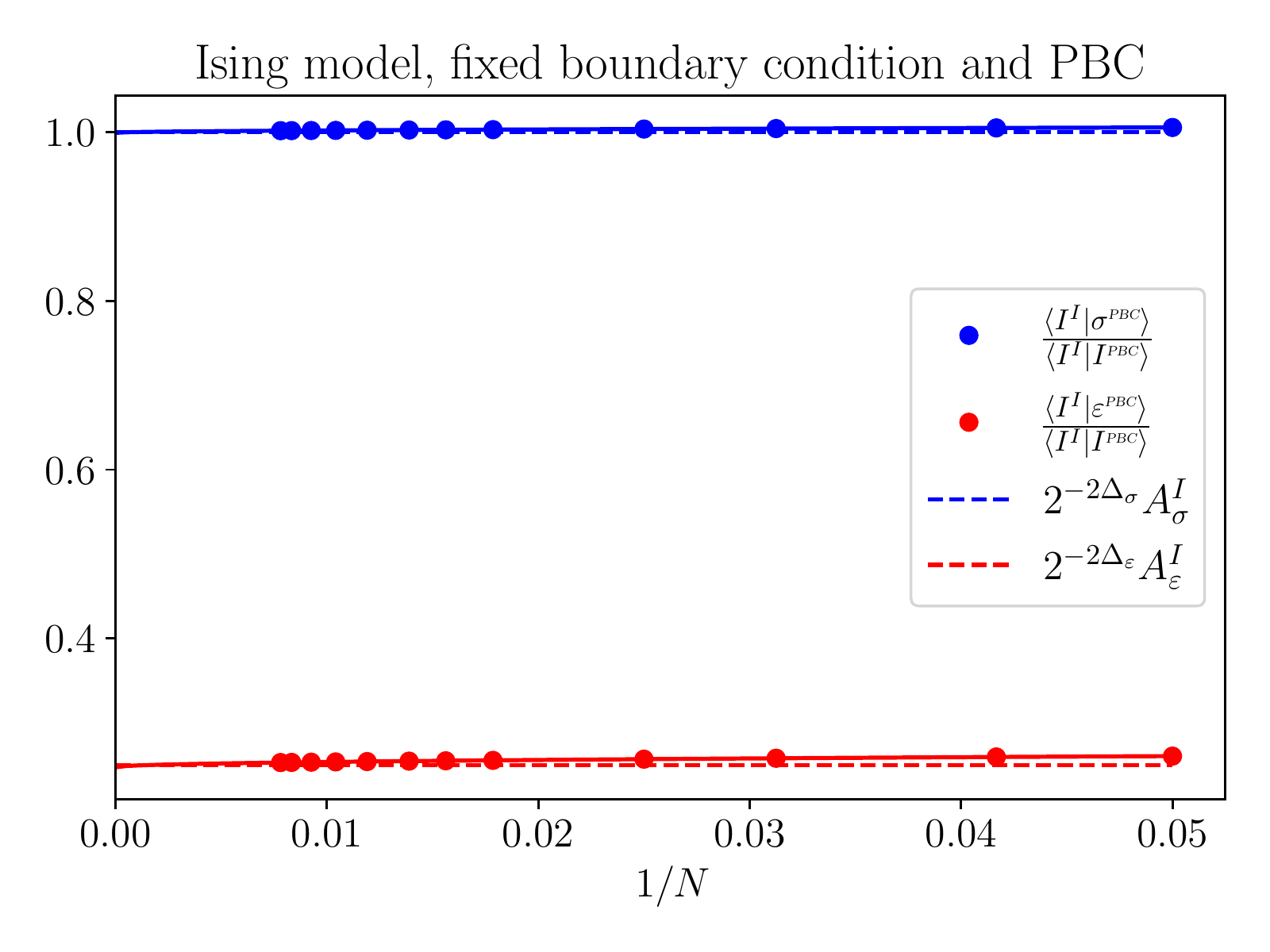}
    \includegraphics[width=0.98\linewidth]{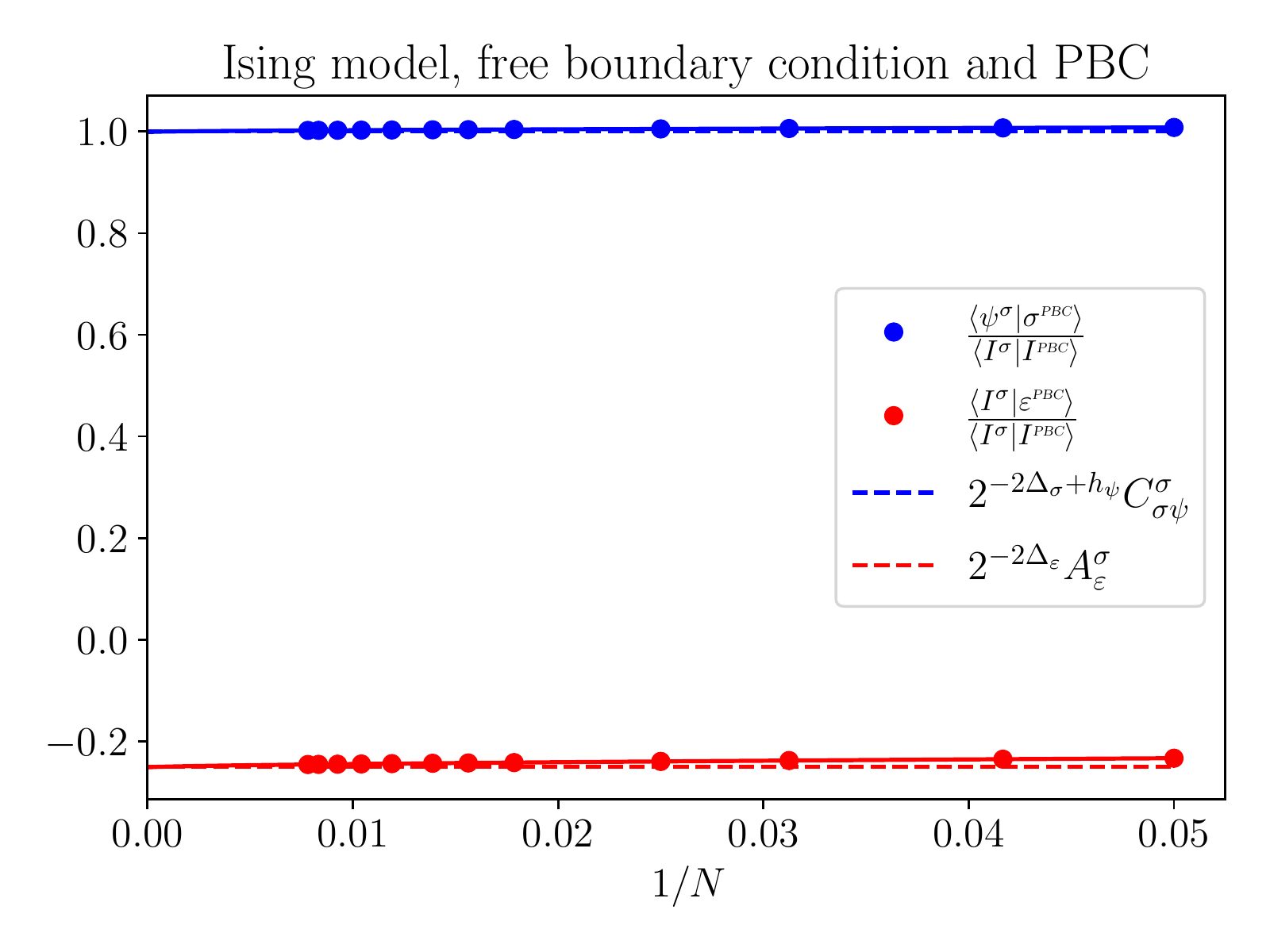}
    \caption{The ratios between overlaps $\langle \psi^{a}_\beta|\phi^{\PBC}_\alpha\rangle$ for the Ising model. The data is obtained by numerical simulations of the Ising model with $20 \leq N \leq 128$. The dashed lines represent the prediction of the main result Eq.~\eqref{eq:main_result2}. The solid lines represent the fitting curve of the finite-size data.}
    \label{fig:Ising2}
\end{figure}

Numerical results of other overlaps are shown in Fig.~\ref{fig:Ising2}. They are also listed in Table~\ref{tab:Ising_OPE} to compare with the theoretical prediction. We see that all numbers agree with the main result Eq.~\eqref{eq:main_result2} to high accuracy. In particular, the minus sign in the bulk-to-boundary OPE can be correctly obtained if the phases of the eigenstates are fixed using the prescription in the last subsection.
\subsection{Three-state Potts model}
The three-state Potts model is described by a unitary minimal CFT with central charge $c=4/5$. The CFT has a $W$ symmetry, which is an extension of the conformal symmetry \cite{fateev_conformal_1987}. With respect to the $W$ symmetry, the bulk primary operators are diagonal combinations of chiral and anti-chiral fields. There are six bulk primary fields, as listed in Table~\ref{tab:Potts_bulk}. The CFT has a $\mathbb{Z}_3$ symmetry, and operators can be labelled with the $\mathbb{Z}_3$ charge. The $\sigma$ and $\sigma^{\dagger}$ fields have the same scaling dimension and opposite charges. They are related by the charge conjugation. The same happens for the $\psi$ and $\psi^{\dagger}$ fields.
\begin{table}[htbp]
    \centering
    \begin{tabular}{|c|c|c|c|}
    \hline
         Bulk operator & $\Delta=h+\bar{h}$ & $s=h-\bar{h}$ & $\mathbb{Z}_3$ charge  \\ \hline
           $I$ &  $0$  & $0$ & $0$\\ \hline
           $\varepsilon$ & $4/5$ & $0$ & $0$\\ \hline
           $\psi$ & $4/3$ & $0$ &$1$\\ \hline
           $\sigma$ & $2/15$ & $0$ & $1$ \\ \hline
           $\psi^{\dagger}$ & $4/3$ & $0$ &$-1$\\ \hline
           $\sigma^{\dagger}$ & $2/15$ & $0$ &$-1$ \\ \hline
    \end{tabular}
    \caption{Bulk primary operators with respect to the $W$ symmetry in the three-state Potts CFT.}
    \label{tab:Potts_bulk}
\end{table}

As with all diagonal rational CFTs, conformal boundary conditions are labelled by the bulk primary fields. For the three-state Potts CFT, the six primary fields label boundary conditions that preserve the $W$ symmetry. The $I, \sigma, \sigma^{\dagger}$ boundary conditions are related by the $\mathbb{Z}_3$ transformations and so are $\varepsilon,\psi,\psi^{\dagger}$ boundary conditions. We will therefore only focus on the $I$ and $\varepsilon$ boundary conditions.

The modular $S$ matrix of the 3-state Potts model is \cite{cardy_boundary_1989}
\begin{equation}
    S = \frac{1}{\sqrt{3}}
    \begin{bmatrix}
     s & s & s \\
     s & \omega s & \omega^2 s \\
     s & \omega^2 s & \omega s
    \end{bmatrix}
\end{equation}
where $\omega = e^{2\pi i/3}$ and
\begin{equation}
    s=\frac{2}{\sqrt{5}}\begin{bmatrix}
    \sin (\pi/5) & \sin(2\pi/5) \\
    \sin (2\pi/5) & -\sin(\pi/5)
    \end{bmatrix}
\end{equation}

We will restrict our attention to the bulk-to-boundary OPE coefficients $A^{a}_{\alpha}\equiv C^{a}_{\alpha I}$, which are determined by the modular $S$ matrix by Eq.~\eqref{eq:A_Smatrix}. Working them out we obtain
\begin{eqnarray}
A^{I}_{\varepsilon}= \varphi, ~~~ A^{I}_{\sigma}= \varphi,~~~ A^{I}_{\psi}= 1, \\
A^{\varepsilon}_{\varepsilon}= -\varphi^{3/2}, A^{\varepsilon}_{\sigma}=-\varphi^{3/2}, A^{\varepsilon}_{\psi}=1,
\end{eqnarray}
where $\varphi = \frac{\sqrt{5}-1}{2}$ is the Golden ratio. 

Now we turn to the lattice model. Let
\begin{equation}
    U= \begin{bmatrix}
    1 & 0 & 0 \\
    0 & \omega & 0 \\
    0 & 0 & \omega^2
    \end{bmatrix},
    V= \begin{bmatrix}
    0 & 1 & 0 \\
    0 & 0 & 1 \\
    1 & 0 & 0
    \end{bmatrix}.
\end{equation}
The Hamiltonian density of the three-state Potts model is 
\begin{equation}
    h^{(2)}_{j,j+1}=-U_j U^{\dagger}_{j+1}- U^{\dagger}_j U_{j+1}, ~~ h^{(1)}_j = -V_j-V^{\dagger}_{j}.
\end{equation}
One lattice realization of the conformal boundary conditions is given by the fixed and mixed boundary conditions \cite{cardy_boundary_1989}. The fixed boundary conditions restrict the boundary spin to a fixed direction, and the mixed boundary conditions restrict the boundary spin to the equal superposition of two directions. They are achieved by the
\begin{eqnarray}
    h^{I}_{L/R} &=& \beta \begin{bmatrix}
    -1 & 0 & 0 \\
    0 & 0 & 0 \\
    0 & 0 & 0
    \end{bmatrix}~~~\, \mathrm{(Fixed)}\\
    h^{\varepsilon}_{L/R} &=& \beta \begin{bmatrix}
    1 & 0 & 0 \\
    0 & 0 & -1 \\
    0 & -1 & 0
    \end{bmatrix} ~\mathrm{(Mixed)}
\end{eqnarray}
where $\beta>0$ is arbitrary. We will use $\beta=8$ in the numerical simulations below. There are two other boundary conditions, including the free boundary condition, that respect the conformal symmetry but not the $W$ symmetry \cite{Affleck_boundary_1998}. We will not consider them here. 

\begin{figure}[hthp]
    \centering
    \includegraphics[width=0.98\linewidth]{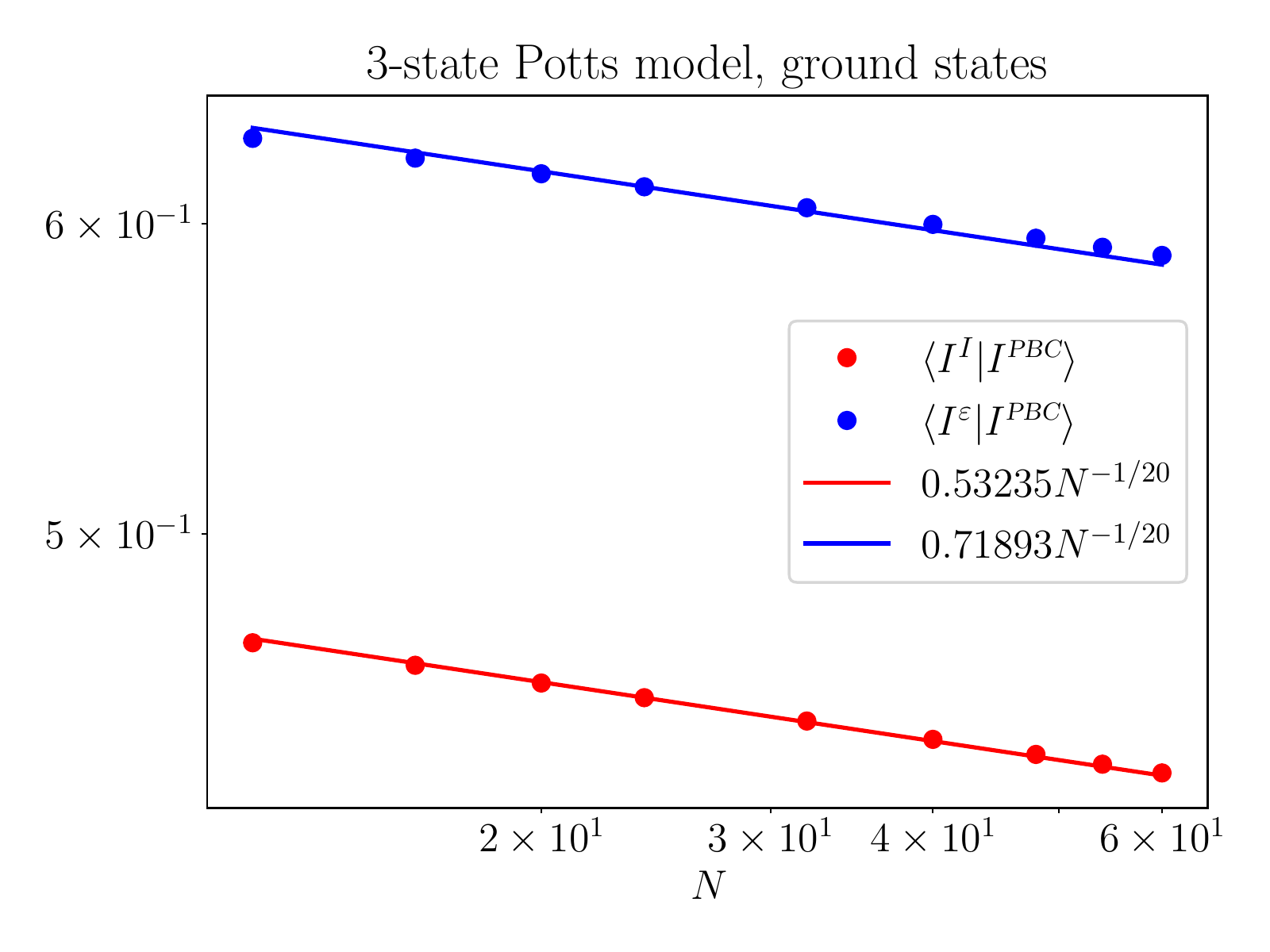}
    \caption{Overlaps between ground states of the three-state model with periodic boundary conditions and open ends. The solid lines represent exponential decaying functions with the correct exponent $-c/16=-1/20$. The data is obtained by numerical simulations of the three-state Potts model with $12 \leq N \leq 60$.}
    \label{fig:Potts1}
\end{figure}

\begin{figure}[htbp]
    \centering
    \includegraphics[width=0.78\linewidth]{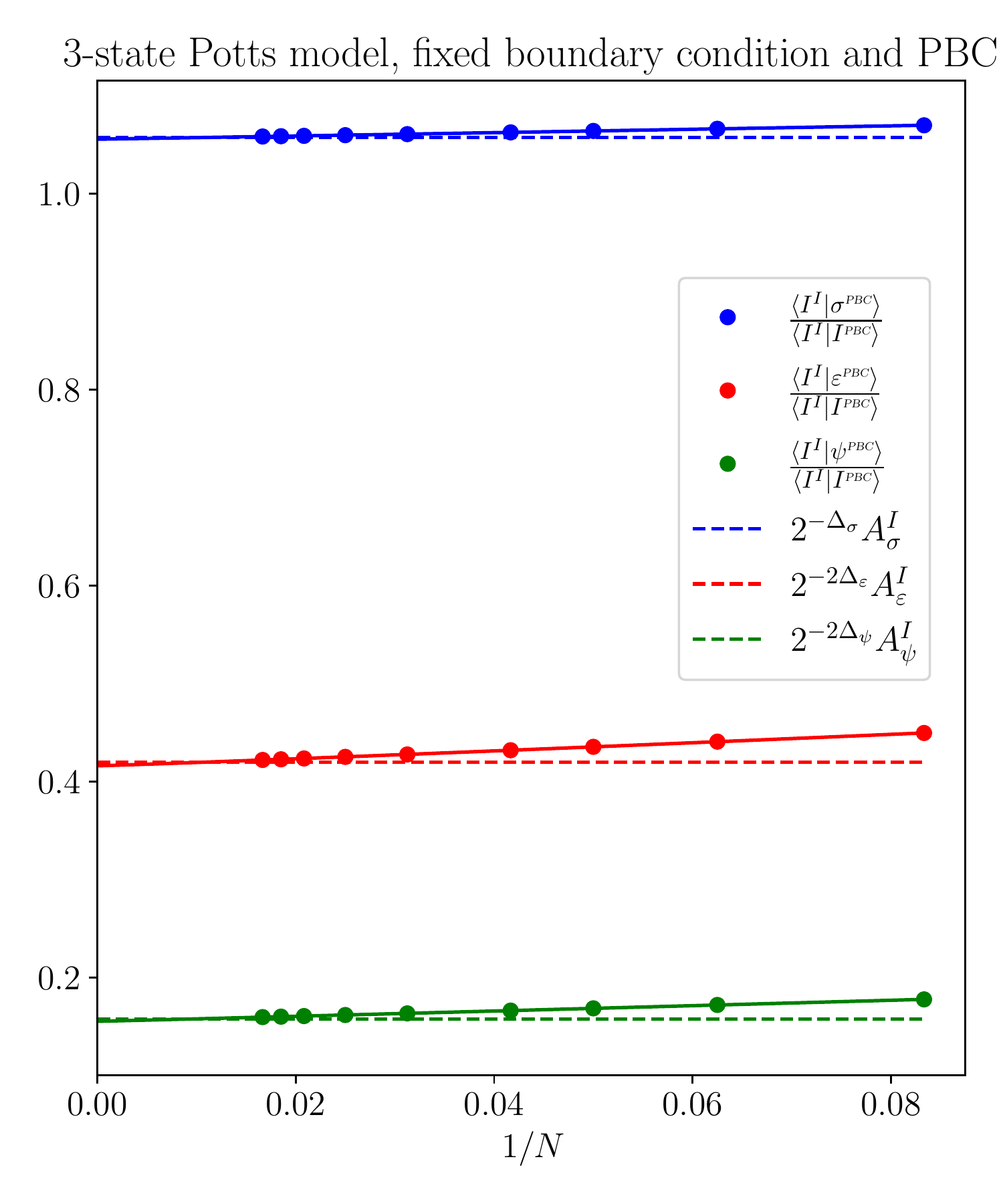}
    \includegraphics[width=0.81\linewidth]{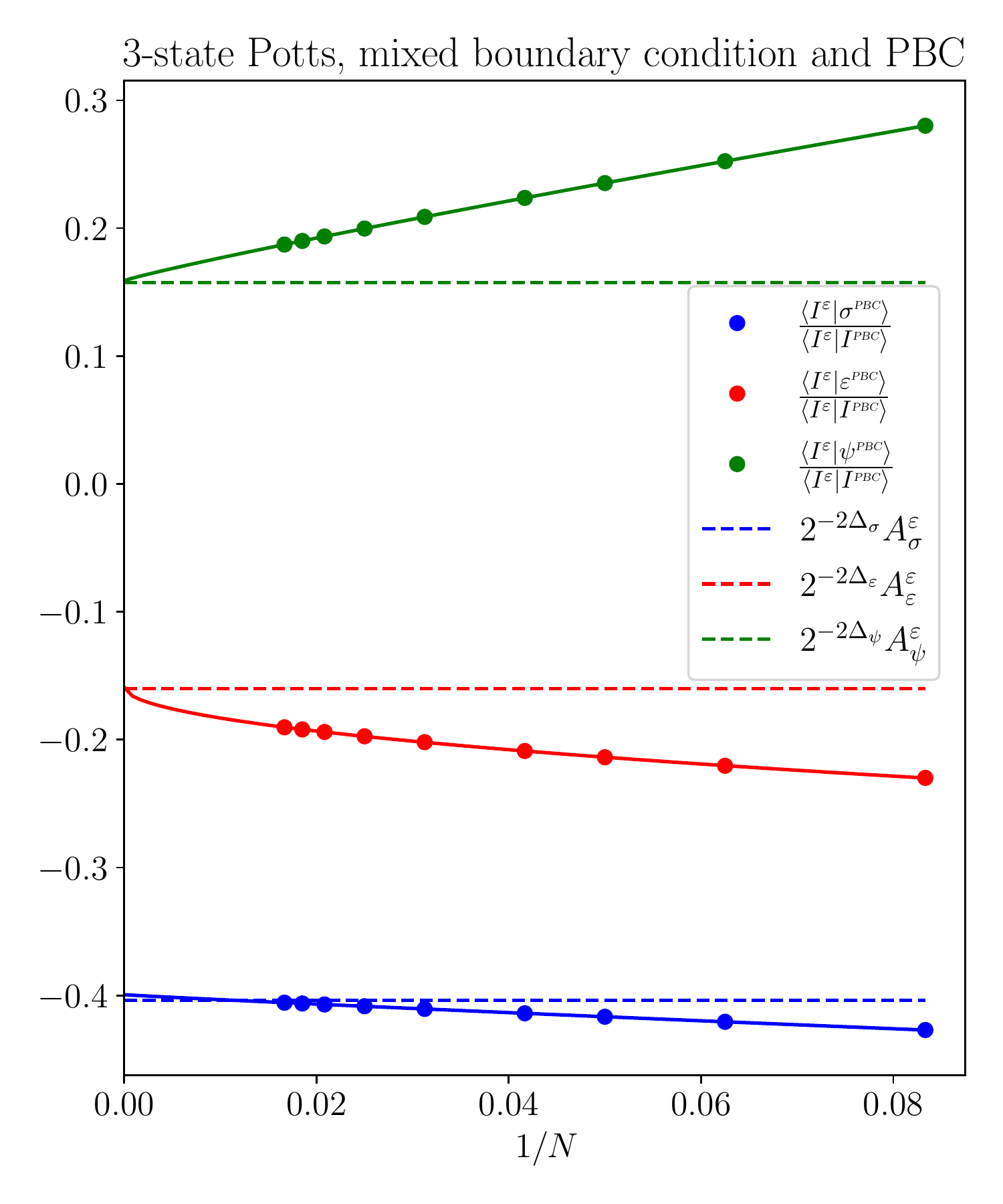}
    \caption{The ratios between overlaps $\langle \psi^{a}_\beta|\phi^{\PBC}_\alpha\rangle$ for the three-state Potts model. The data is obtained by numerical simulations of the Ising model with $12 \leq N \leq 60$. The dashed lines represent the prediction of the main result Eq.~\eqref{eq:main_result2}. The solid lines represent the fitting curve of the finite-size data.}
    \label{fig:Potts2}
\end{figure}

The overlap between ground states decays as
\begin{equation}
    \langle I^{a}|I^{\PBC}\rangle = \mathcal{N}_a N^{-1/20},
\end{equation}
which we plot in Fig.~\ref{fig:Potts1}. The overlaps that correspond to the bulk-to-boundary OPEs are listed in Table~\ref{tab:Potts_OPE}.  
\begin{table}[htbp]
    \centering
    \begin{tabular}{|c|c|c|c|}
    \hline
            $\phi^{\PBC}_{\alpha}$ & $\psi^{a}_{\beta}$ & \makecell{$\bra{\psi^{a}_{\beta}} \phi^{\PBC}_{\alpha}\rangle/\langle I^{a}| I^{\PBC}\rangle$ \\ (Numerical value)} & \makecell{$2^{-2\Delta_\alpha+h_{\beta}} C^{a}_{\alpha\beta}$ \\ (CFT data)} \\ \hline
           $\sigma$ & $I^{I}$ & 1.056&  1.0574 \\ \hline
           $\sigma$ & $I^{\varepsilon}$ &$-0.399$& $-0.4039$\\ \hline
           $\varepsilon$ & $I^{I}$ & 0.416& $0.4196$\\ \hline
           $\varepsilon$ & $I^{\varepsilon}$ & $-0.159$& $-0.1603$ \\ \hline
           $\psi$ & $I^{I}$ &0.155 & $0.1575$ \\ \hline
           $\psi$ & $I^{\varepsilon}$ &0.159 & $0.1575$ \\ \hline
    \end{tabular}
    \caption{The ratios of the overlaps for the three-state Potts model. The first two columns specify the two states in the overlap, and the third column shows the ratio $\bra{\psi^{a}_{\beta}} \phi^{\PBC}_{\alpha}\rangle/\langle I^{a}| I^{\PBC}\rangle$ extrapolated to the thermodynamic limit (LHS of Eq.~\eqref{eq:main_result1}). The fourth column shows the corresponding CFT quantity $2^{-2\Delta_\alpha+h_{\beta}} C^{a}_{\alpha\beta}$ (RHS of Eq.~\eqref{eq:main_result2}). For the spin chain with open ends, only the ground state is considered.}
    \label{tab:Potts_OPE}
\end{table}
The extrapolation of finite-size data is shown in Fig.~\ref{fig:Potts2}. Again we see excellent agreement with the theoretical prediction Eq.~\eqref{eq:main_result2}.
\section{Discussion}
\label{sect:conclusion}
In this work we have studied the overlaps $\langle \psi^{a}_{\beta}|\phi^{\PBC}_{\alpha}\rangle$ between the low-energy eigenstates of a critical quantum spin chain, where $|\phi^{\PBC}_{\alpha}\rangle$ is an eigenstate of the spin chain with PBC and $|\psi^{a}_{\beta}\rangle$ is an eigenstate of the spin chain with boundary condition $a$.  The main result Eqs.~\eqref{eq:main_result1} and \eqref{eq:main_result2} indicates that the universal information characterizing the CFT, including the central charge $c$ and the bulk-to-boundary OPE coefficients $C^{a}_{\alpha\beta}$, can be extracted from the overlaps. 

For a generic critical quantum spin chain, it is often hard to identify the lattice operators with CFT operators, thus limiting our ability to extract the conformal data from correlation functions. Our result is useful because it is purely based on the low-energy eigenstates as opposed to correlation functions. Conceptually, our result indicates that the low-energy eigenstates of a critical quantum spin chain already encode universal information about the phase transition. The key insight is to use the operator-state correspondence to translate the expressions in terms of operators into the expressions in terms of states. 

In this work we have restricted our attention to the overlaps between eigenstates of the spin chain with periodic boundary conditions or open ends. This is analogous to the closed-open string amplitude in string theory \cite{lewellen_sewing_1992}. It is natural to consider the geometries that correspond to more general string amplitudes that involve several closed or open strings. This will allow us to compute higher-point correlation functions involving bulk and boundary operators using overlaps between eigenstates. We will explore in this direction in future work.

\textit{Acknowledgements} -- The author is grateful to Guifre Vidal for helpful discussions. The author is particularly in debt to Qi Hu for initializing a related project. Sandbox is a team within the Alphabet family of companies, which includes Google, Verily, Waymo, X, and others. The DMRG calculations in this work are performed with ITensor library \cite{itensor}.

\appendix
\section{Computing overlaps using matrix product states}
In order to reduce finite-size corrections, we use the matrix product state (MPS) method to compute the low-energy eigenstates of the critial quantum spin chain.

For spin chains with open ends, the low-energy eigenstates of $H^{a}$ can be obtained by DMRG. The state is represented as a MPS of the following form,
\begin{equation}
    |\psi\rangle=M^{s_1}_{1}\cdots M^{s_N}_{N} \ket{\vec{s}},
\end{equation}
where $M^{s_n}_n$ is a matrix with shape $(\chi^{(n)}_l, \chi^{(n)}_r)$, and $\chi^{(n)}_l=\chi^{(n-1)}_r, \chi^{(n)}_r=\chi^{(n+1)}_l$ are bond dimensions. Note that $\chi^{(1)}_l=\chi^{(N)}_r=1$. We will use $\chi$ to denote the largest bond dimensions among all bonds. The numerical cost of the DMRG algorithm is $O(N\chi^3)$.

For the spin chain with PBC, we use the periodic uniform MPS method to diagonalize the low-energy eigenstates with definite momenta. The eigenstates are of the following form,
\begin{equation}
    |\phi\rangle=\sum_{j=1}^{N} e^{ipj}\mathrm{Tr}(A^{s_1}\cdots B^{s_j}\cdots A^{s_N}) \ket{\vec{s}},
\end{equation}
where $A,B$ are tensors with shape $(D,d,D)$, and $p$ is the momentum. The dominant numerical cost of the puMPS algorithm is $O(ND^5)$.

Finally, the overlap $\langle \psi|\phi\rangle$ can be computed as follows. Let $T^{A}_M$ denote the mixed transfer matrix
\begin{equation}
    (T^A_M)_{(\alpha\gamma)(\beta\delta)} \equiv \sum_s A^{s}_{\alpha\beta}\bar{M}^{s}_{\gamma\delta},
\end{equation}
which is regarded as a matrix with row index $(\alpha\gamma)$ and column index $(\beta\delta)$. 
Then 
\begin{equation}
    \langle \psi|\phi\rangle = \sum_{j=1}^{N} e^{ipj} \mathrm{Tr}(T^{A}_{M_1}\cdots T^{B}_{M_j} \cdots T^{A}_{M_N}).
\end{equation}
We can compute the sum efficiently with the algorithm below.
\begin{eqnarray}
    &~&I_A=T^A_{M_1}; I_B=T^B_{M_1} e^{ip}; \nonumber\\
    &~&\mathbf{for}~j~\mathrm{in}~2:N \nonumber\\
    &~& ~~~~I_B = I_BT^{A}_{M_j}+I_AT^{B}_{M_j} e^{ipj}; \nonumber\\ 
    &~& ~~~~I_A = I_AT^{A}_{M_j}; \nonumber\\
    &~&\mathbf{end} \nonumber \\
    &~& \langle \psi |\phi \rangle = \sum_{\alpha\gamma }(I_B)_{\alpha\gamma\alpha\gamma}; \nonumber
\end{eqnarray}
The numerical cost to compute the overlap is $O(N(\chi D^3+\chi^2 D^2))$. In practice this is significantly smaller than the puMPS ground-state optimization which takes $O(ND^5)$.
 
\bibliography{puMPS_CFT}
\end{document}